\documentclass{article}

\usepackage[nonatbib, preprint]{neurips_2022}

\usepackage[utf8]{inputenc} 
\usepackage[T1]{fontenc}    
\usepackage{hyperref}       
\usepackage{url}            
\usepackage{booktabs}       
\usepackage{amsfonts}       
\usepackage{nicefrac}       
\usepackage{microtype}      
\usepackage{xcolor}         
\usepackage{comment}

\usepackage[numbers]{natbib}

\usepackage{amsmath}
\usepackage{graphicx}

\title{Towards a non-Gaussian Generative Model of large-scale Reionization Maps}

\author{%
Yu-Heng Lin   \\
  Department of Physics\\
  University of Minnesota\\
  Minneapolis, MN 55455 \\
  \texttt{lin00025@umn.edu} \\
\And 
Sultan Hassan \\
  Center for Computational Astrophysics\\
  Flatiron Institute\\
  New York, NY 10010 \\ 
  Department of Astrophysical Sciences\\
  Princeton University, Peyton Hall\\
  Princeton, NJ, 08544\\
  Department of Physics \& Astronomy \\
  University of the Western Cape \\
  Cape Town 7535\\
  \texttt{shassan@flatironinstitute.org} \\ NHFP Hubble Fellow \\
\And 
Bruno Régaldo-Saint Blancard   \\
  Center for Computational Mathematics\\
  Flatiron Institute\\
  New York, NY 10010 \\
  \texttt{bregaldosaintblancard@flatironinstitute.org} \\
\And 
Michael Eickenberg   \\
  Center for Computational Mathematics\\
  Flatiron Institute\\
  New York, NY 10010 \\
  \texttt{meickenberg@flatironinstitute.org} \\
\And
Chirag Modi   \\
  Center for Computational Astrophysics\\
  Flatiron Institute\\
  New York, NY 10010 \\
  \texttt{cmodi@flatironinstitute.org} \\
}

\begin{document}

\maketitle

\begin{abstract}
High-dimensional data sets are expected from the next generation of large-scale surveys. These data sets will carry a wealth of information about the early stages of galaxy formation and cosmic reionization. Extracting the maximum amount of information from the these data sets remains a key challenge. Current simulations of cosmic reionization are computationally too expensive to provide enough realizations to enable testing different statistical methods, such as parameter inference. We present a non-Gaussian generative model of reionization maps that is based solely on their summary statistics. We reconstruct large-scale ionization fields (bubble spatial distributions) directly from their power spectra (PS) and Wavelet Phase Harmonics  (WPH) coefficients. Using WPH, we show that our model is efficient in generating diverse new examples of large-scale ionization maps from a single realization of a summary statistic. We compare our model with the target ionization maps using the bubble size statistics, and largely find a good agreement. As compared to PS, our results show that WPH provide optimal summary statistics that capture most of information out of a highly non-linear ionization fields.      
\end{abstract}

\section{Introduction}
In the early universe, neutral gas accreted to the over-density region inside dark matter halos, and formed the first generations of galaxies.  The intergalactic medium was later ionized by the ultraviolet photons emitted from these galaxies.  Studying this epoch, known as the cosmic reionization, is crucial to understand the earliest stages of galaxy formation and evolution.  Various properties of high-redshift galaxies are poorly constrained and cannot be measured directly. Instead, studying the integrated emission from these galaxies over large-scales, a technique known as intensity mapping, is emerging as a powerful cosmological probe. Several intensity mapping experiments, such as SKA \citep{SKA}, HERA \citep{DeBoer_2017}, LOFAR \citep{van_Haarlem_2013}, Euclid \citep{Euclid_Collaboration_2020}, SPHEREx \citep{SPHEREx}, and Roman \citep{WFIRST_2015}, are expected to provide large-scale maps in different bands, including hydrogen ionization maps in the early universe.

Translating these growing observational efforts into astrophysical and cosmological constraints on our theoretical models of reionization and galaxy formation remains a key challenge. One limitation is the computational cost of reionization simulations, which is an obstacle to generate enough samples of detailed large-scale maps, fully explore the parameter space controlling different astrophysical ingredients, and perform parameter inference. However, many of these experiments focus on statistical measurements  (e.g. the power spectrum). Efficient sampling from summary statistics is therefore required in order to extract most of information from the upcoming reionization surveys.

In this work,  we introduce a non-Gaussian generative model of large-scale ionization maps that is based on wavelet phase harmonic (WPH) statistics \citep{Allys_2020, PyWPH_2021, Regaldo2022}. We compare it to a Gaussian model constrained by power spectrum statistics.  We then use the bubble-size statistics as an independent metric to compare the input map with our reconstructed maps.

\section{Methods}\label{sec:method}

Simulations of cosmic reionization generate ionization fields using the following steps: (i) generation and evolution of the initial density field. (ii) identification of the sources (galaxies/halos) with group finder methods. (iii) computation of the radiative transfer to generate the ionization field using the density and source fields at different epochs. Steps (ii) and (iii) are computationally very expensive, thus we aim to accelerate them with our fast generative model.

\begin{figure*}[h]
  \centering
  \includegraphics[width=0.8\columnwidth]{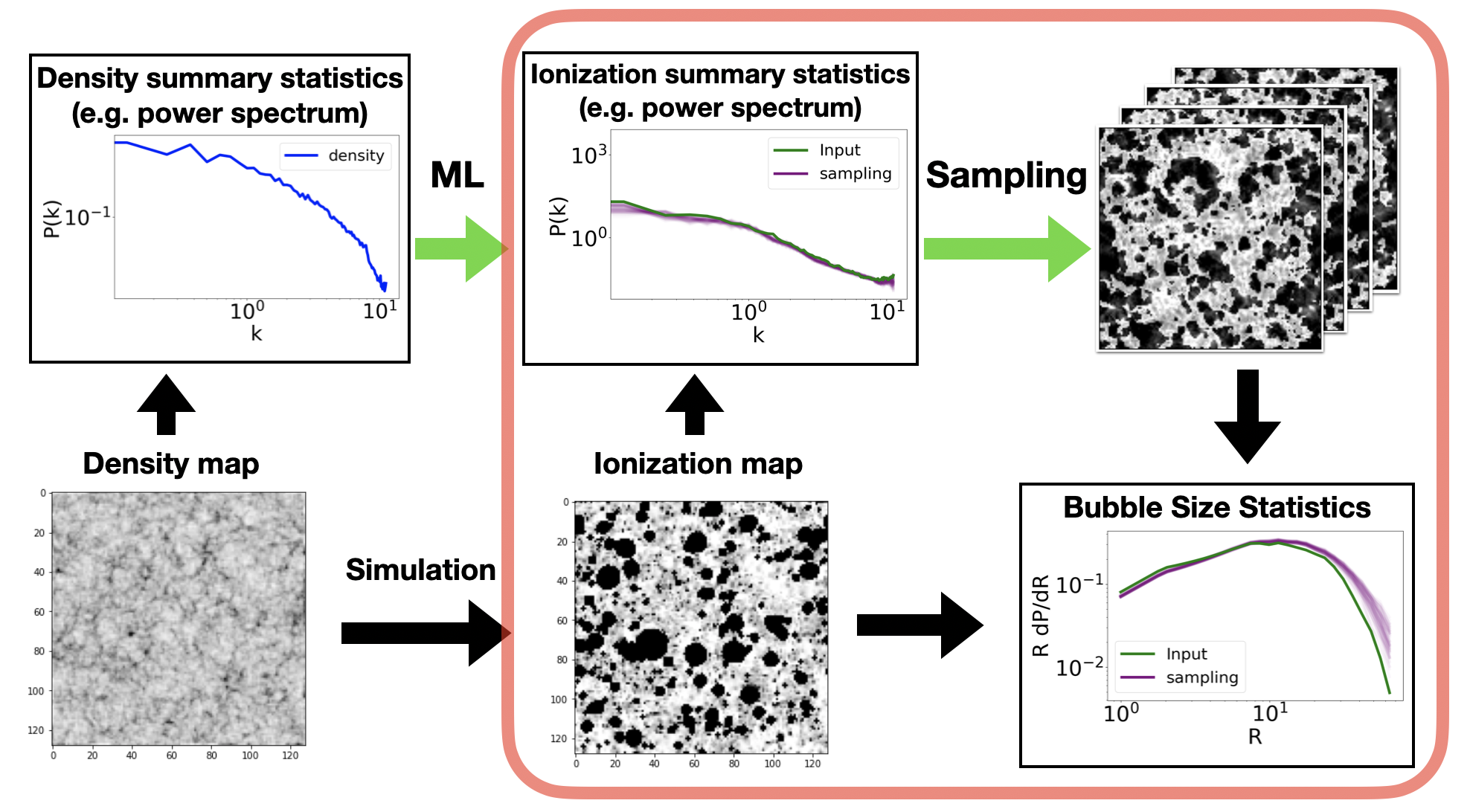}
  \caption{The pipeline of the generative model. This forward generative model (green arrows) transfers the density summary statistics to the ionization summary statistics and then reconstructs the ionization field. We test the reconstructed maps using the bubble size statistics against the target map. This pipeline consists of two parts and the red box indicates the focus of this paper.}
  \label{fig:pipeline}
\end{figure*}
Our paper focuses on the sampling procedure of the following proposed pipeline:
\begin{itemize}
    \item Create conditional mapping between the summary statistics of the density field and the summary statistics of the ionization field, using either Multilayer perceptron or Symbolic regression (This is beyond the scope of this paper).
    \item Reconstruct the ionization field from their summary statistics (This is the focus of this paper).
\end{itemize}
We present a visual summary of the pipeline in Figure~\ref{fig:pipeline}.  To explore the quality of the reconstructed maps, we use different summary statistics, and compare the results using the bubble size statistics.

To create the input ionization field, we use the semi-numerical simulations SimFast21 \citep{Santos_2010}. We choose the physical scale length of the field as 250 Mpc, and the field size as 128$\times$128$\times$128 voxels. As mentioned earlier, the simulation first generates a density field, identifies the ionizing sources from the overdensity region, then creates the ionization field. Using SimFast21, we generate 100 reionization realizations and each simulation run takes approximately 5 minutes.

\subsection{Generative model}\label{}

In this section, we build generative models that are adapted from~\cite{Regaldo2022}, and we make use of the GPU-accelerated public Python package PyWPH~\cite{PyWPH_2021} to compute WPH statistics.
In general, wavelet transforms extract information by convolving the input with a series of kernels/filters of different orientations and scales, and then integrate the resulting convolved input after applying the absolute magnitude operator as a non-linear activation function \citep{mallat1999wavelet}. In our case, the filters are defined using phase harmonics. One can think of wavelets transform as a convolutional neural network without training since the kernels are fixed to predefined wavelets. 

 Let $\phi_i$ be the operator that computes the $i$th summary statistics from a map. 
Samples of these models are generated by drawing a 128$\times$128 white noise map $u_0$ and iteratively deforming it so that the statistics of this map match those of the target $s$ (in our case, the ionization map generated from SimFast21). This is done by minimizing the following loss function: 
\begin{equation}
L_i(u) =  | \phi_i(u) - \phi_i(s) |^2 . 
\end{equation}
The minimization is performed using an LBFGS optimizer \citep{Byrd_1995}.  Additional details on this approach can be found in \cite{Regaldo2022}.

We refer to the generative model that relies on the $\phi_i$ operator that computes the power spectrum as PS model and to WPH statistics as WPH model.
The WPH model computes a large number of coefficients since it captures the higher order statistics (e.g. non-Gaussinity). For the purpose of building a forward model, we explore different combinations of WPH moments to reduce the size of coefficients. In this work, we show the model that computes the S$^{(1,1)}$ moments and the scaling moments L, denoted as S$^{(1,1)}$+L model. 
The S$^{(1,1)}$ moment contains similar information to the power spectrum, and the scaling moments L relate to one-point statistics. 
A detailed description of the WPH moments can be found in \cite{Allys_2020, Regaldo2022}.

\begin{figure*}[t]
  \centering
  \includegraphics[scale=0.4]{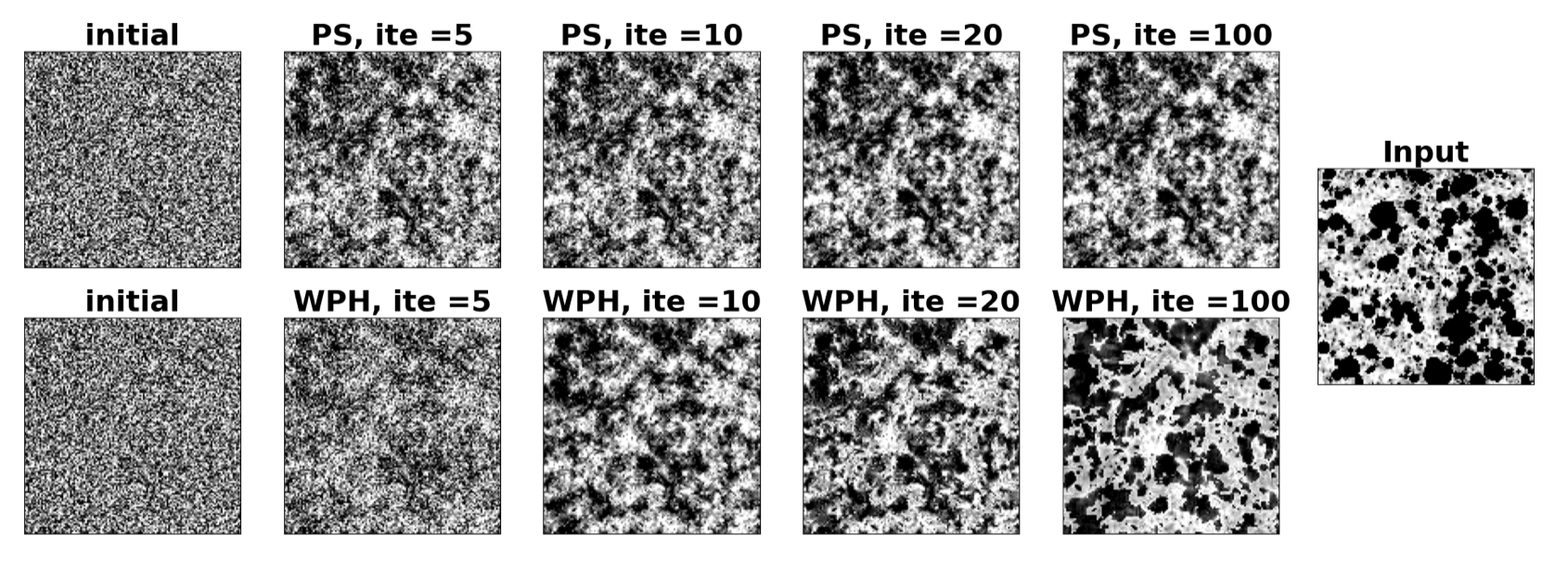}
  \caption{ Sampling using the PS model (upper) and the WPH model (lower) at different iterations. }
  \label{fig:iter}
\end{figure*}

In Figure~\ref{fig:iter}, we show how the PS and WPH models deform the white noise over iteration. The PS model quickly produces patchy ionized region, but has no significant improvement after 20 iterations.  The WPH model shows very different outcome after 20 iterations.     
For each model, we run the optimization up to 500 iterations. 
The calculations are done using 1 GPU v100-16 on the PSC Bridge-2 cluster \citep{XSEDE}.  
Each iteration takes 0.06 seconds.  It takes roughly 30 seconds to generate a realization. 
While the PS model has a smaller size of coefficients,  it takes roughly 10 seconds to generate a realization. 
Our model is faster by a factor of 10-30 as compared to the benchmark simulation.
The pixel values of the generated images vary between -1 to 2, as a consequence of optimization.  Since the neutral fraction can only be between 0 and 1, we clip the values of all the realizations at 0 and 1 after the optimization. We generate 100 realizations for each model.

\subsection{Bubble Size Statistics}

To measure the ionization bubble size,  we use the mean-free-path method from Tool21cm\footnote{\url{https://tools21cm.readthedocs.io/}}\citep{Giri_2018}.
This method was introduced in \cite{Mesinger_2007}, and was deeply discussed in \cite{Friedrich_2011, Lin_2016, Giri_2018}. It first converts the ionization map into a binary field based on whether the ionized fraction x$_{\rm HII}$ of a location is above or below a threshold value. We choose 0.7 as our threshold.  A random direction would be selected out of a random ionized spot to casts out a photon. The photon vanishes after hitting the first neutral point, and the traveling distance is recorded.  We repeat this process 10$^6$ times and report the frequency of these traveling distances, and refer to the histogram as the bubble-size distribution (BSD).

\begin{figure*}[t]
  \centering
  \includegraphics[width=0.8\columnwidth]{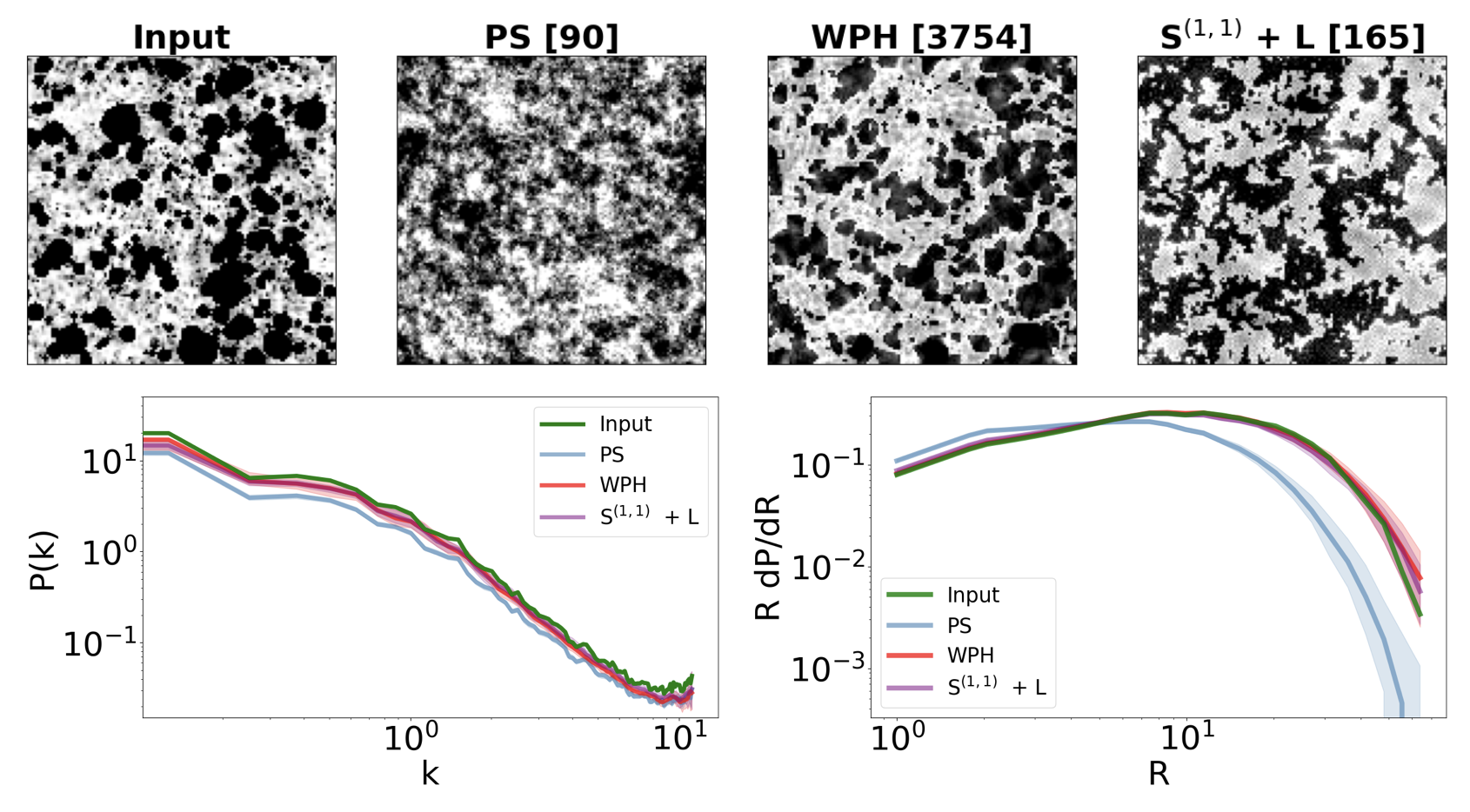}
  \caption{Upper panel: The input ionization image, images generated with power spectrum, WPH statistics, and the S$^{(1,1)}$ + L moments.  The numbers of the coefficients are reported in the brackets. Lower panel: The power spectra and bubble statistics of the images, with 5$\%$ and 95$\%$ percentiles shown as the shaded area.  }
  \label{fig:result}
\end{figure*}

\section{Results}

In this section we compare our results to the input ionization map. On the upper panel of Figure~\ref{fig:result}, we show the input ionization image and one sample of each model.  On the lower panel,  we show the mean power spectra and BSD of realizations,with 5$\%$ and 95$\%$ percentiles shown as the shaded area.

The PS model generates images with fuzzy edges since the power spectrum does not capture the non-Gaussian features. 
The WPH model, on the other hand, is able to capture the non-Gaussianity and successfully recreates the rounded shape of ionization bubbles. However, the WPH model produces more realistic outputs due to the large number of coefficients (3754), that capture more information (e.g. non-gaussinity), as opposed to the power spectrum coefficients (90).  The S$^{(1,1)}$+L model reduces the size of the coefficients to 165, and the generated images show clear separation of neutral and ionized regions. 

All models recover the input power spectrum decently well. For each generative model, the scatter of the power spectra between different realizations are mostly less than a factor of 2.  The outputs of the the PS model before clipping have an identical power spectrum to the input image, but drop by 70$\%$ after clipping the values between 0 and 1. The clipping does not affect the outputs from the WPH models, since the resulting pixel values are already close to the expected dynamic range of 0 and 1.

The bubble size statistic describes not only the bubble radius but also the percolation of the reionization.  
The WPH models perform much better than the PS model in BSD, especially at producing large-scale bubbles. 
In the mean-free path method, the images are converted to a binary field based on a threshold neutral fraction, which we choose x$_{\rm HII}$=0.7 as the threshold.  Unlike the case with the target and generated maps from WPH model, the PS models do not generate sharper edges of ionized regions. Instead, these PS models  create smoother transformations between the neutral and the ionized region.  Therefore, the BSD of the PS model is strongly dependent on the choice of the threshold. In summary, our results suggest that WPH models successfully produce synthetic large scale ionization maps whose power spectra and BSD are similar to the target.

\section{Conclusion}
We have presented a fast non-Gaussian generative model of large-scale reionization maps that is based solely on summary statistics. We have used different statistics: the power spectrum statistics and the WPH statistics. We find that the latter generates more realistic synthetic ionization maps as compared to the former. This is due to the fact that the WPH statistics capture much more information (non-Gaussianity) than the power spectrum. Our generative models are faster than the benchmark simulations by a factor of 10-30, which can accelerate statistical analysis, such as parameter inference. We are currently working on the first part of the pipeline to create a conditional mapping between the WPH statistics of the density field and those of the ionization field on several astrophysical parameters such as the photon escape fraction. We are testing different methods such as Multilayer perceptron or Symbolic regression to create the conditional mapping to complete the design of our fast non-Gaussian generative model.

\section{Broader Impact}
Next-generation of large-scale surveys are expected to deliver large amount of big data sets which requires a new generation of tools/models to enable extracting most of information. We presented a fast generative model that would be useful for testing statistical analysis pipelines and forecasting to experiments that rely on measuring summary statistics. While this method is applied on reionization maps, a similar method could be used in different fields of astrophysics and cosmology to accelerate costly simulations. All fields of physics will be able to benefit from this work whenever the aim is to generate highly non-linear large scale fields. 



\bibliographystyle{unsrt}

\bibliography{mybib}

\begin{thebibliography}{10}

\bibitem{SKA}
Ska observatory.
\newblock \url{https://www.skatelescope.org/}.

\bibitem{DeBoer_2017}
David~R. DeBoer, Aaron~R. Parsons, James~E. Aguirre, Paul Alexander, Zaki~S.
  Ali, Adam~P. Beardsley, Gianni Bernardi, Judd~D. Bowman, Richard~F. Bradley,
  Chris~L. Carilli, Carina Cheng, Eloy de~Lera~Acedo, Joshua~S. Dillon, Aaron
  Ewall-Wice, Gcobisa Fadana, Nicolas Fagnoni, Randall Fritz, Steve~R.
  Furlanetto, Brian Glendenning, Bradley Greig, Jasper Grobbelaar, Bryna~J.
  Hazelton, Jacqueline~N. Hewitt, Jack Hickish, Daniel~C. Jacobs, Austin
  Julius, MacCalvin Kariseb, Saul~A. Kohn, Telalo Lekalake, Adrian Liu, Anita
  Loots, David MacMahon, Lourence Malan, Cresshim Malgas, Matthys Maree,
  Zachary Martinot, Nathan Mathison, Eunice Matsetela, Andrei Mesinger,
  Miguel~F. Morales, Abraham~R. Neben, Nipanjana Patra, Samantha Pieterse,
  Jonathan~C. Pober, Nima Razavi-Ghods, Jon Ringuette, James Robnett, Kathryn
  Rosie, Raddwine Sell, Craig Smith, Angelo Syce, Max Tegmark, Nithyanandan
  Thyagarajan, Peter K.~G. Williams, and Haoxuan Zheng.
\newblock Hydrogen epoch of reionization array ({HERA}).
\newblock {\em Publications of the Astronomical Society of the Pacific},
  129(974):045001, mar 2017.

\bibitem{van_Haarlem_2013}
{van Haarlem, M. P.}, {Wise, M. W.}, {Gunst, A. W.}, {Heald, G.}, {McKean, J.
  P.}, {Hessels, J. W. T.}, {de Bruyn, A. G.}, {Nijboer, R.}, {Swinbank, J.},
  {Fallows, R.}, {Brentjens, M.}, {Nelles, A.}, {Beck, R.}, {Falcke, H.},
  {Fender, R.}, {H\"orandel, J.}, {Koopmans, L. V. E.}, {Mann, G.}, {Miley,
  G.}, {R\"ottgering, H.}, {Stappers, B. W.}, {Wijers, R. A. M. J.}, {Zaroubi,
  S.}, {van den Akker, M.}, {Alexov, A.}, {Anderson, J.}, {Anderson, K.}, {van
  Ardenne, A.}, {Arts, M.}, {Asgekar, A.}, {Avruch, I. M.}, {Batejat, F.},
  {B\"ahren, L.}, {Bell, M. E.}, {Bell, M. R.}, {van Bemmel, I.}, {Bennema,
  P.}, {Bentum, M. J.}, {Bernardi, G.}, {Best, P.}, {B\^{\i}rzan, L.},
  {Bonafede, A.}, {Boonstra, A.-J.}, {Braun, R.}, {Bregman, J.}, {Breitling,
  F.}, {van de Brink, R. H.}, {Broderick, J.}, {Broekema, P. C.}, {Brouw, W.
  N.}, {Br\"uggen, M.}, {Butcher, H. R.}, {van Cappellen, W.}, {Ciardi, B.},
  {Coenen, T.}, {Conway, J.}, {Coolen, A.}, {Corstanje, A.}, {Damstra, S.},
  {Davies, O.}, {Deller, A. T.}, {Dettmar, R.-J.}, {van Diepen, G.}, {Dijkstra,
  K.}, {Donker, P.}, {Doorduin, A.}, {Dromer, J.}, {Drost, M.}, {van Duin, A.},
  {Eisl\"offel, J.}, {van Enst, J.}, {Ferrari, C.}, {Frieswijk, W.}, {Gankema,
  H.}, {Garrett, M. A.}, {de Gasperin, F.}, {Gerbers, M.}, {de Geus, E.},
  {Grie\ss{}meier, J.-M.}, {Grit, T.}, {Gruppen, P.}, {Hamaker, J. P.},
  {Hassall, T.}, {Hoeft, M.}, {Holties, H. A.}, {Horneffer, A.}, {van der
  Horst, A.}, {van Houwelingen, A.}, {Huijgen, A.}, {Iacobelli, M.}, {Intema,
  H.}, {Jackson, N.}, {Jelic, V.}, {de Jong, A.}, {Juette, E.}, {Kant, D.},
  {Karastergiou, A.}, {Koers, A.}, {Kollen, H.}, {Kondratiev, V. I.},
  {Kooistra, E.}, {Koopman, Y.}, {Koster, A.}, {Kuniyoshi, M.}, {Kramer, M.},
  {Kuper, G.}, {Lambropoulos, P.}, {Law, C.}, {van Leeuwen, J.}, {Lemaitre,
  J.}, {Loose, M.}, {Maat, P.}, {Macario, G.}, {Markoff, S.}, {Masters, J.},
  {McFadden, R. A.}, {McKay-Bukowski, D.}, {Meijering, H.}, {Meulman, H.},
  {Mevius, M.}, {Middelberg, E.}, {Millenaar, R.}, {Miller-Jones, J. C. A.},
  {Mohan, R. N.}, {Mol, J. D.}, {Morawietz, J.}, {Morganti, R.}, {Mulcahy, D.
  D.}, {Mulder, E.}, {Munk, H.}, {Nieuwenhuis, L.}, {van Nieuwpoort, R.},
  {Noordam, J. E.}, {Norden, M.}, {Noutsos, A.}, {Offringa, A. R.}, {Olofsson,
  H.}, {Omar, A.}, {Orr\'u, E.}, {Overeem, R.}, {Paas, H.}, {Pandey-Pommier,
  M.}, {Pandey, V. N.}, {Pizzo, R.}, {Polatidis, A.}, {Rafferty, D.},
  {Rawlings, S.}, {Reich, W.}, {de Reijer, J.-P.}, {Reitsma, J.}, {Renting, G.
  A.}, {Riemers, P.}, {Rol, E.}, {Romein, J. W.}, {Roosjen, J.}, {Ruiter, M.},
  {Scaife, A.}, {van der Schaaf, K.}, {Scheers, B.}, {Schellart, P.},
  {Schoenmakers, A.}, {Schoonderbeek, G.}, {Serylak, M.}, {Shulevski, A.},
  {Sluman, J.}, {Smirnov, O.}, {Sobey, C.}, {Spreeuw, H.}, {Steinmetz, M.},
  {Sterks, C. G. M.}, {Stiepel, H.-J.}, {Stuurwold, K.}, {Tagger, M.}, {Tang,
  Y.}, {Tasse, C.}, {Thomas, I.}, {Thoudam, S.}, {Toribio, M. C.}, {van der
  Tol, B.}, {Usov, O.}, {van Veelen, M.}, {van der Veen, A.-J.}, {ter Veen,
  S.}, {Verbiest, J. P. W.}, {Vermeulen, R.}, {Vermaas, N.}, {Vocks, C.},
  {Vogt, C.}, {de Vos, M.}, {van der Wal, E.}, {van Weeren, R.}, {Weggemans,
  H.}, {Weltevrede, P.}, {White, S.}, {Wijnholds, S. J.}, {Wilhelmsson, T.},
  {Wucknitz, O.}, {Yatawatta, S.}, {Zarka, P.}, {Zensus, A.}, and {van Zwieten,
  J.}
\newblock Lofar: The low-frequency array.
\newblock {\em A\&A}, 556:A2, 2013.

\bibitem{Euclid_Collaboration_2020}
{Euclid Collaboration}, A.~{Blanchard}, S.~{Camera}, C.~{Carbone}, V.~F.
  {Cardone}, S.~{Casas}, S.~{Clesse}, S.~{Ili{\'c}}, M.~{Kilbinger},
  T.~{Kitching}, M.~{Kunz}, F.~{Lacasa}, E.~{Linder}, E.~{Majerotto},
  K.~{Markovi{\v{c}}}, M.~{Martinelli}, V.~{Pettorino}, A.~{Pourtsidou},
  Z.~{Sakr}, A.~G. {S{\'a}nchez}, D.~{Sapone}, I.~{Tutusaus},
  S.~{Yahia-Cherif}, V.~{Yankelevich}, S.~{Andreon}, H.~{Aussel},
  A.~{Balaguera-Antol{\'\i}nez}, M.~{Baldi}, S.~{Bardelli}, R.~{Bender},
  A.~{Biviano}, D.~{Bonino}, A.~{Boucaud}, E.~{Bozzo}, E.~{Branchini},
  S.~{Brau-Nogue}, M.~{Brescia}, J.~{Brinchmann}, C.~{Burigana}, R.~{Cabanac},
  V.~{Capobianco}, A.~{Cappi}, J.~{Carretero}, C.~S. {Carvalho}, R.~{Casas},
  F.~J. {Castander}, M.~{Castellano}, S.~{Cavuoti}, A.~{Cimatti},
  R.~{Cledassou}, C.~{Colodro-Conde}, G.~{Congedo}, C.~J. {Conselice},
  L.~{Conversi}, Y.~{Copin}, L.~{Corcione}, J.~{Coupon}, H.~M. {Courtois},
  M.~{Cropper}, A.~{Da Silva}, S.~{de la Torre}, D.~{Di Ferdinando},
  F.~{Dubath}, F.~{Ducret}, C.~A.~J. {Duncan}, X.~{Dupac}, S.~{Dusini},
  G.~{Fabbian}, M.~{Fabricius}, S.~{Farrens}, P.~{Fosalba}, S.~{Fotopoulou},
  N.~{Fourmanoit}, M.~{Frailis}, E.~{Franceschi}, P.~{Franzetti}, M.~{Fumana},
  S.~{Galeotta}, W.~{Gillard}, B.~{Gillis}, C.~{Giocoli},
  P.~{G{\'o}mez-Alvarez}, J.~{Graci{\'a}-Carpio}, F.~{Grupp}, L.~{Guzzo},
  H.~{Hoekstra}, F.~{Hormuth}, H.~{Israel}, K.~{Jahnke}, E.~{Keihanen},
  S.~{Kermiche}, C.~C. {Kirkpatrick}, R.~{Kohley}, B.~{Kubik},
  H.~{Kurki-Suonio}, S.~{Ligori}, P.~B. {Lilje}, I.~{Lloro}, D.~{Maino},
  E.~{Maiorano}, O.~{Marggraf}, N.~{Martinet}, F.~{Marulli}, R.~{Massey},
  E.~{Medinaceli}, S.~{Mei}, Y.~{Mellier}, B.~{Metcalf}, J.~J. {Metge},
  G.~{Meylan}, M.~{Moresco}, L.~{Moscardini}, E.~{Munari}, R.~C. {Nichol},
  S.~{Niemi}, A.~A. {Nucita}, C.~{Padilla}, S.~{Paltani}, F.~{Pasian}, W.~J.
  {Percival}, S.~{Pires}, G.~{Polenta}, M.~{Poncet}, L.~{Pozzetti}, G.~D.
  {Racca}, F.~{Raison}, A.~{Renzi}, J.~{Rhodes}, E.~{Romelli}, M.~{Roncarelli},
  E.~{Rossetti}, R.~{Saglia}, P.~{Schneider}, V.~{Scottez}, A.~{Secroun},
  G.~{Sirri}, L.~{Stanco}, J.~L. {Starck}, F.~{Sureau},
  P.~{Tallada-Cresp{\'\i}}, D.~{Tavagnacco}, A.~N. {Taylor}, M.~{Tenti},
  I.~{Tereno}, R.~{Toledo-Moreo}, F.~{Torradeflot}, L.~{Valenziano},
  T.~{Vassallo}, G.~A. {Verdoes Kleijn}, M.~{Viel}, Y.~{Wang}, A.~{Zacchei},
  J.~{Zoubian}, and E.~{Zucca}.
\newblock {Euclid preparation. VII. Forecast validation for Euclid cosmological
  probes}.
\newblock {\em A\&A}, 642:A191, October 2020.

\bibitem{SPHEREx}
Spherex.
\newblock \url{https://spherex.caltech.edu/}.

\bibitem{WFIRST_2015}
D.~{Spergel}, N.~{Gehrels}, C.~{Baltay}, D.~{Bennett}, J.~{Breckinridge},
  M.~{Donahue}, A.~{Dressler}, B.~S. {Gaudi}, T.~{Greene}, O.~{Guyon},
  C.~{Hirata}, J.~{Kalirai}, N.~J. {Kasdin}, B.~{Macintosh}, W.~{Moos},
  S.~{Perlmutter}, M.~{Postman}, B.~{Rauscher}, J.~{Rhodes}, Y.~{Wang},
  D.~{Weinberg}, D.~{Benford}, M.~{Hudson}, W.~S. {Jeong}, Y.~{Mellier},
  W.~{Traub}, T.~{Yamada}, P.~{Capak}, J.~{Colbert}, D.~{Masters}, M.~{Penny},
  D.~{Savransky}, D.~{Stern}, N.~{Zimmerman}, R.~{Barry}, L.~{Bartusek},
  K.~{Carpenter}, E.~{Cheng}, D.~{Content}, F.~{Dekens}, R.~{Demers},
  K.~{Grady}, C.~{Jackson}, G.~{Kuan}, J.~{Kruk}, M.~{Melton}, B.~{Nemati},
  B.~{Parvin}, I.~{Poberezhskiy}, C.~{Peddie}, J.~{Ruffa}, J.~K. {Wallace},
  A.~{Whipple}, E.~{Wollack}, and F.~{Zhao}.
\newblock {Wide-Field InfrarRed Survey Telescope-Astrophysics Focused Telescope
  Assets WFIRST-AFTA 2015 Report}.
\newblock {\em arXiv e-prints}, page arXiv:1503.03757, March 2015.

\bibitem{Allys_2020}
E.~Allys, T.~Marchand, J.-F. Cardoso, F.~Villaescusa-Navarro, S.~Ho, and
  S.~Mallat.
\newblock New interpretable statistics for large-scale structure analysis and
  generation.
\newblock {\em Phys. Rev. D}, 102:103506, Nov 2020.

\bibitem{PyWPH_2021}
{Regaldo-Saint Blancard, Bruno}, {Allys, Erwan}, {Boulanger, Fran\c{c}ois},
  {Levrier, Fran\c{c}ois}, and {Jeffrey, Niall}.
\newblock A new approach for the statistical denoising of planck interstellar
  dust polarization data.
\newblock {\em A\&A}, 649:L18, 2021.

\bibitem{Regaldo2022}
Bruno {R{\'e}galdo-Saint Blancard}, Erwan {Allys}, Constant {Auclair},
  Fran{\c{c}}ois {Boulanger}, Michael {Eickenberg}, Fran{\c{c}}ois {Levrier},
  L{\'e}o {Vacher}, and Sixin {Zhang}.
\newblock {Generative Models of Multi-channel Data from a Single Example --
  Application to Dust Emission}.
\newblock {\em arXiv e-prints}, page arXiv:2208.03538, August 2022.

\bibitem{Santos_2010}
M.~G. Santos, L.~Ferramacho, M.~B. Silva, A.~Amblard, and A.~Cooray.
\newblock {Fast large volume simulations of the 21-cm signal from the
  reionization and pre-reionization epochs}.
\newblock {\em Monthly Notices of the Royal Astronomical Society},
  406(4):2421--2432, 08 2010.

\bibitem{mallat1999wavelet}
St{\'e}phane Mallat.
\newblock {\em A wavelet tour of signal processing}.
\newblock Elsevier, 1999.

\bibitem{Byrd_1995}
Richard~H. Byrd, Peihuang Lu, Jorge Nocedal, and Ciyou Zhu.
\newblock A limited memory algorithm for bound constrained optimization.
\newblock {\em SIAM Journal on Scientific Computing}, 16(5):1190--1208, 1995.

\bibitem{XSEDE}
J.~Towns, T.~Cockerill, M.~Dahan, I.~Foster, K.~Gaither, A.~Grimshaw,
  V.~Hazlewood, S.~Lathrop, D.~Lifka, G.~D. Peterson, R.~Roskies, J.~Scott, and
  N.~Wilkins-Diehr.
\newblock Xsede: Accelerating scientific discovery.
\newblock {\em Computing in Science `I\&' Engineering}, 16(05):62--74, sep
  2014.

\bibitem{Giri_2018}
Sambit~K. Giri, Garrelt Mellema, Keri~L. Dixon, and Ilian~T. Iliev.
\newblock {Bubble size statistics during reionization from 21-cm tomography}.
\newblock {\em Monthly Notices of the Royal Astronomical Society},
  473(3):2949--2964, 10 2017.

\bibitem{Mesinger_2007}
Andrei Mesinger and Steven Furlanetto.
\newblock Efficient simulations of early structure formation and reionization.
\newblock {\em The Astrophysical Journal}, 669(2):663--675, nov 2007.

\bibitem{Friedrich_2011}
Martina~M. {Friedrich}, Garrelt {Mellema}, Marcelo~A. {Alvarez}, Paul~R.
  {Shapiro}, and Ilian~T. {Iliev}.
\newblock {Topology and sizes of H II regions during cosmic reionization}.
\newblock {\em Monthly Notices of the Royal Astronomical Society},
  413(2):1353--1372, May 2011.

\bibitem{Lin_2016}
Yin Lin, S.~Peng Oh, Steven~R. Furlanetto, and P.~M. Sutter.
\newblock {The distribution of bubble sizes during reionization}.
\newblock {\em Monthly Notices of the Royal Astronomical Society},
  461(3):3361--3374, 06 2016.

\end{thebibliography}

\end{document}